# Taking off with Biodegradable Tensegrities: An Eco-friendly Emergency Medical Delivery Solution

Madhumati Anand
Amrita Vidyalayam Puthiyakavu
Karunagapally, Kerala
India
http://pkvu.amritavidyalayam.org
madhumati.anand@gmail.com

Vyzag Ajith
Amrita Vidyalayam Puthiyakavu
Karunagapally, Kerala
India
http://pkvu.amritavidyalayam.org
vyzagajith2k19@gmail.com

Sanjula Sreekumar
Amrita Vidyalayam Puthiyakavu
Karunagapally, Kerala
India
http://pkvu.amritavidyalayam.org
sanjula2453@gmail.com

## ABSTRACT
This paper analyzes how medical supplies can be safely transported through eco-friendly means to areas affected by disasters such as floods using Unmanned Aerial Vehicles. We compared the performance of different bio-degradable tensegrity structures with fragile payloads cushioned with coir padding by conducting drop tests from successively greater heights upto 75 meters on surfaces with different hardness. Our results showed that the bio-degradable tensegrities was able to absorb the shock on impact successively in all cases without any damage to the payload. The results suggest that biodegradable tensegrities are a viable option for fragile payload delivery.

## Keywords
Tensegrities; humanitarian technologies; experimental approach; disaster relief.

## 1. INTRODUCTION

Today, Unmanned Aerial Vehicles (UAVs) provide an option for delivering critical supplies quickly to those who are temporarily isolated and marooned during natural disasters such as floods [1]. UAVs are robots that fly and can be controlled remotely or autonomously by a program. UAVs are of several types including fixed-wing planes, copters and hybrids. They are also classified as small, medium, larger, larger, etc. [2] according to their Maximum Gross Takeoff Weight (MGTW). We are using a small UAV for our project with Vertical Take Off and Landing (VTOL) because these UAVs can be built and flown by amateurs and is ideal for carrying small payloads (upto 500 gms).

Now, is there a way to provide emergency relief using low-cost eco-friendly materials using small UAVs to those who are inaccessible through land routes? Tensegrity is one such a model that can be used for air-dropping fragile payloads [3, 4]. Tensegrity is a structure which consists of two components – struts (sticks) and strings (cords) which are in a state of constant compression and tension respectively. Since tensegrities are extremely flexible and compressible structures, they can be used to cushion the fall of a fragile payload by absorbing and dissipating the kinetic energy upon impact. The ball-like shape of a tensegrity allows it to bounce and roll which further helps in shock absorption.

## 2. PROJECT DESCRIPTION

### 2.1 Project Overview

This work presents an idea for medical/emergency relief payload transportation using a small quadcopter so as to reach out to maximum number of people in a disaster-affected area. The capacity of the payload ranges from 300 gm to 500 gm supply of vital medicines such as insulin, antibiotics, and serum vials that cannot be simply air-dropped without rupture and breakage. The structure we are presenting consists of the medicine payload wrapped in a multi-layered coir fiber blanket. This bundle is tied in the center of a tensegrity. When this tensegrity contraption is dropped from a quadcopter, due to its unique construction, it will compress and roll, thereby dissipating the kinetic energy upon ground impact. The coir blanket around the payload further protects the payload from breaking.

To keep this tensegrity contraption eco-friendly, all the components of the contraption (other than the payload itself) are made of bio-degradable materials. The coir blanket is made up of recycled coir from used coconuts; the tensegrities are made up of wicker cane/bamboo sticks and jute and wool strings. The tensegrity contraption is made with locally available bio-degradable materials because the contraption is primarily meant for two to three drops. After fulfilling its purpose of payload delivery of emergency relief supplies, it can be disposed of.

Early experimentation of the tensegrity contraption with a single tensegrity structure helped us understand that it could successfully break the fall of a fragile payload from smaller heights such as 20 feet and 30 feet. But when the same tensegrity structure was dropped from 80 feet height, it could not absorb enough kinetic energy upon impact to prevent the payload from breaking. Hence we enhanced the tensegrity structure to add the coir padding for payload protection. Also, our original tensegrity used elastic strings. We replaced the elastic strings with non-elastic jute strings which gave the entire structure enough rigidity to not compress upon impact with the ground.

The emergency medical relief supplies we chose to test with included insulin vials, injection needles and tablets. We chose this payload because Diabetes Mellitus (DM) is a very common problem for many people in the state of Kerala, India [5, 6]. Research reports put the prevalence of DM in Kerala at nearly 20% [7] thereby earning the state the dubious distinction of being the diabetes capital of India. We spoke to several first responders from the 2017 Kerala floods and learned that the maximum medical supply requests came for insulin injections. If diabetes patients do not take their insulin medications on time, it could lead to strokes, cardiac problems, blindness and permanent nerve damage. For all these reasons, we chose to perform our drop tests of the tensegrity structure with a payload of diabetes medicine supply that would be sufficient for two to four persons for a day.

## 2.2 Drop test experiments

We performed a series of drop tests from a drone with the insulin medicine payload, starting from 25 meters to successively greater heights up to 75 meters. We tested the fall on a soft ground and on a hard pavement with a small payload of 2 insulin glass vials, 10 injection needles and 1 strip of 10 tablets and a larger payload twice the quantity (4 insulin glass vials, 10 injection needles and 20 tablets). Each time, our payload stayed intact without breaking. However, after four to five drops, the jute strings of the tensegrity became slack. This made the tensegrity wobbly and we needed to rebuild a new tensegrity to continue experimenting. The video at https://youtu.be/2OiAO3vpUJ8 captures our work so far.

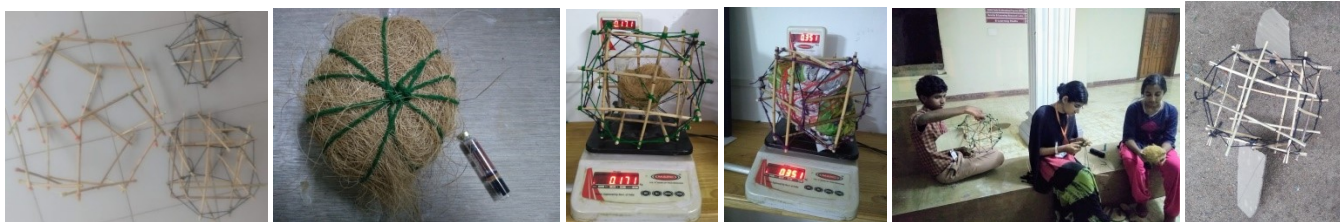

| 6 strut, 18 strut and 30 sturt tensegrities | Insulin medicines payload wrapped in coir | Different payload weights that we performed drop tests with | Division of work between the three of us | Tensegrity with flaps |

## 2.3 Lessons Learned

This project has been a wonderful learning experience for all of us. Though we are classmates and see each other every day in school, this was our first STEM project together as a team. We learned how to divide up the tasks, do research, keep systematic documentation of our learnings from each experiment we performed. Among the many lessons learned, we also learned that that there are many things that will not work but if we put our heart into it, the aspirations will get realized. Since the project has evolved slowly over 9 months, we had the opportunity to learn several lessons during tensegrity construction itself:

- Tensegrities built with cane sticks and jute strings are extremely lightweight. Also the jute strings are non-elastic, so the tensegrities can be built into a firm shape. But, unlike the elastic string tensegrities which 'spring' back into shape when unpacked, the non-elastic tensegrities have to be taken apart for packing and remade upon unpacking.
- Since the jute strings are non-elastic, it requires much more precision than the elastic string tensegrities. The jute strings have to be cut to the right length. To know the correct length of the jute strings to build the tensegrity into a firm structure, it is helpful to make a tensegrity with elastic strings first, and measure the final string length in the built tensegrity before rebuilding with jute.
- The jute strings have very high tensile strength. In order to make the tensegrity, we made wedges on both ends of the cane sticks and slipped the jute string through it. We had to tie the mouth of the wedges because if we didn't, the jute strings would just slice the cane sticks because of the tension they were in. When that happens, the tensegrity would lose its shape and just collapse.
- Since jute is a natural fiber, the jute strings gets frayed after a couple of drop tests. When that happens, the strings become weak and can snap. Hence the biodegradable tensegrities cannot be reused more than 4-5 times without rebuilding them.

We observed that different tensegrities have different features:

- The 6 strut, 24 string icosahedron tensegrity is the smallest and easiest to build but its interior space allows only for a very small payload for the same strut length as other tensegrity designs. Its structure allows for rolling and is very robust against failures.
- The 18 strut tensegrity cube is a very modular design where each side of the cube (shaped like an H) can be pre-made and stacked. Should a disaster strike, the tensegrity cube can be quickly assembled in less than 15 minutes. Its interior volume allows for payloads up to 4 insulin vials along with 10 injections and 20 tablets surrounded by a thick layer of coir padding.
- The 30 strut, 30 string dodecahedron tensegrity is most spherical in shape though it is relatively harder to assemble and build. It lacks inner structural strength since there are no cross-members but it has large interior volume for protecting the payload.

Our work is not complete - it is a work in progress. We are continuing to perform drop tests with a 30 strut tensegrity and also improve the design by adding flaps to increase the drag to slow down the descent.

## 3. BIOS

Madhumati Anand is a 9[th] grade student in Amrita Vidyalayam, Puthiyakavu. She loves to read books, paint, create and play with innovative scientific toys and learns Indian classical and semi-classical music and dance. She will be a member of the panel.

Sanjula Sreekumar is a 9[th] grader in Amrita Vidyalayam, Puthiyakavu. She loves to dance, choreograph dances and explore science through experimentation. She aspires to become a doctor and serve her country through her selfless service. She will be a member of the panel.

Vyzag Ajith is a 9[th] grade student in Amrita Vidyalayam, Puthiyakavu. He is a science enthusiast and an aspiring astrophysicist. He loves music and plays the violin. He is a pet lover, numismatist and also likes to play football. He will be a member of the panel.